\documentclass{sig-alternate-10pt}

\usepackage{epsfig}
\usepackage{url}
\usepackage{endnotes}
\usepackage{graphicx,amsmath,amssymb,subfigure,url,xspace,color}
\usepackage{color}
\usepackage{breakurl}
\usepackage{multirow}
\usepackage{tabularx}
\usepackage{threeparttable}
\usepackage{flushend}

\newcommand{\eg}{e.g.,\xspace}

\newcommand{\etc}{etc.\@\xspace}
\newcommand{\ie}{i.e.,\xspace}

\newcommand{\unit}[1]{\ensuremath{\, \mathrm{#1}}}

\graphicspath{{./figures/}}

\begin{document}
\title{\Large \bf Low-Power Distance Bounding}
\date{}
\numberofauthors{1}
\author{\alignauthor Aanjhan Ranganathan, Boris Danev, Srdjan Capkun\\
 \affaddr{Institute of Information Security} \\
 \affaddr{Dept. of Computer Science, ETH Zurich} \\
 \affaddr{Zurich, Switzerland } \\
 \email{raanjhan@inf.ethz.ch, boris.danev@inf.ethz.ch, capkuns@inf.ethz.ch}
}

\maketitle

\begin{abstract}

  A distance bounding system guarantees an upper bound on the physical distance
  between a verifier and a prover. However, in contrast to a conventional
  wireless communication system, distance bounding systems introduce tight
  requirements on the processing delay at the prover and require high distance
  measurement precision making their practical realization challenging. Prior
  proposals of distance bounding systems focused primarily on building provers
  with minimal processing delays but did not consider the power limitations of
  provers and verifiers. However, in a wide range of applications (e.g.,
  physical access control), provers are expected to be fully or semi passive
  introducing additional constraints on the design and implementation of
  distance bounding systems.

  In this work, we propose a new physical layer scheme for distance bounding and
  leverage this scheme to implement a distance bounding system with a low-power
  prover. Our physical layer combines frequency modulated continuous wave (FMCW)
  and backscatter communication. The use of backscatter communication enables
  low power consumption at the prover which is critical for a number of distance
  bounding applications. By using the FMCW-based physical layer, we further
  decouple the physical distance estimation from the processing delay at the
  prover, thereby enabling the realization of the majority of distance bounding
  protocols developed in prior art. We evaluate our system under various attack
  scenarios and show that it offers strong security guarantees against distance,
  mafia and terrorist frauds. Additionally, we validate the communication and
  distance measurement characteristics of our system through simulations and
  experiments and show that it is well suited for short-range physical access
  control and payment applications.
\end{abstract}

\section{Introduction}
\label{sec:introduction}
The widespread deployment of wireless systems that use location and
proximity to provide services has led to the advent of many radio frequency
based localization technologies~\cite{LiuNov07}. Today, these systems are used
in a broad set of scenarios including people and asset tracking, emergency and
rescue support~\cite{fischer10} and access
control~\cite{GuptaMar06,RasmussenNov09}. Given the safety and security
implications of the above mentioned applications, it is important to ensure the
security of the location estimate and data used in these systems.

Distance bounding enables the secure measurement of an upper bound on the
physical distance between two devices, a verifier and a prover, even if the
prover is untrusted and tried to reduce the measured distance. Distance bounding
was initially introduced in the context of wired systems~\cite{BrandsMay93} and
later a number of distance bounding
protocols~\cite{TippenhauerSep09,TuSep07,HanckeSep05,RasmussenAug10,MunillaNov08,ReidMar07,BussardMay05,CapkunOct03,Lopez09,Lopez10,SingeleeJul07}
were designed for wireless systems. In-order to compute the upper bound on the
physical distance, distance bounding relies on the measurement of the round-trip
time between a transmitted challenge and a received response. Successful
execution of a distance bounding protocol relies on two main assumptions: (i)
Precise distance bound estimate and (ii) Low processing time at the prover to
compute the response. Precise measurement of the distance depends largely on the
physical characteristics of the RF signal and the time-of-arrival estimation
technique implemented in the system. The time taken by the prover to process the
challenge (\ie demodulate, compute and transmit the response) depends on the
chosen processing function and is critical to prevent distance modification
attacks such as distance fraud~\cite{BrandsMay93} or mafia
fraud~\cite{DesmedtAug87}. Reducing this processing time is therefore critical,
such that the prover cannot modify its processing time arbitrarily and pretend
to be closer to the verifier. Some prior work on prover design focused on using
analog or hybrid digital-analog processing in order to reduce the prover
processing time to few nanoseconds~\cite{RasmussenAug10,RanganathanSep12}. Those
designs focus primarily on the prover architecture without much consideration
for the physical characteristics (modulation scheme, bandwidth, encoding, bit
periods etc.) of the radio communication signals, which form a critical part of
a distance bounding system.

Another line of work considered the implementation of distance bounding using
ultra-wide band (UWB) signals with well defined physical-layer characteristics.
Tippenhauer~\cite{Tippenhauer12} implemented a distance bounding system with a
prover processing delay of approximately $100\unit{ns}$. This limits the
distance modification by an untrusted prover to maximum $15\unit{m}$. While the
proposed distance bounding implementation was well specified, it is not clear
whether the prover processing time can be reduced to a few nanoseconds and
whether such prover designs can be made practical for power sensitive
applications (e.g., RFID localization, proximity-based electronic tokens for
access control and mobile payments).

The realization of low power provers for distance bou\-nding is very important
for the development of practical distance bounding systems in many of today's
applications. For example, RFID technology is used in a number of applications
ranging from identification and tracking of commodity goods, physical access
control, animal husbandry tracking, automatic toll collection systems,
electronic passports and payment systems. Prior research has revealed that the
use of RFID proximity to provide access control is vulnerable to mafia-fraud
(relay) attacks (\eg PKES systems~\cite{FrancillonFeb11}, NFC
phones~\cite{FrancisDec10}, Google Wallet~\cite{RolandSep12}). The ability to
realize distance bounding protocols for passive or semi-passive RFID devices (or
tags) would prevent the majority of relay attack scenarios.

Our physical layer scheme uses the frequency modulated continuous wave (FMCW)
for distance estimation and On-Off Keying technique for data communication. We
show that due to the inherent nature of FMCW, the distance estimation phase is
only loosely coupled to the challenge processing at the prover \ie the distance
estimation is independent of the processing delay at the prover while keeping
the security guarantees of the system intact. This enables logical layer
implementation of any distance bounding protocol proposed in prior art. Our
proposed system architecture offers complete protection against distance fraud
attacks where a dishonest, but trusted prover tries to cheat on the distance by
processing the challenges faster. An attacker does not gain any distance
advantage by replying earlier or processing the challenges faster. In addition,
we provide maximum distance reduction estimates for a strong attacker who is
capable of detecting challenges earlier and relaying them to a trusted prover.

In this work, we propose a new distance bounding system designed for
short-range, low-power applications. Specifically, we make the following
contributions.
\begin{itemize}
\item We propose and evaluate a new physical layer sche\-me specifically designed
  for the realization of distance bounding systems.
\item Leveraging this physical layer scheme, we design a prover that can
  potentially be integrated into passive and semi-passive RFID tags, thus
  enabling distance bounding for power constrained applications.
\item We analyze our system under distance, mafia and terrorist fraud attacks
  and show how our system resists these attacks.
\item We evaluate our system through simulations and experimentally validate its
  processing delay, power consumption and ranging precision.
\end{itemize}

The remainder of the paper is organized as follows. In
Section~\ref{sec:db-background} we introduce distance bounding and briefly
discuss the existing distance bounding systems and their current limitations. In
Section~\ref{sec:fmcw-based-db}, we provide the essentials of FMCW and describe
our physical layer scheme for distance bounding. We analyze our system against
the known distance, mafia and terrorist fraud attacks in
Section~\ref{sec:security-analysis} and experimentally evaluate the design in
Section~\ref{sec:system-evaluation}. Finally, we discuss future work and
conclude the paper in Section~\ref{sec:conclusion}.

\section{Distance Bounding}
\label{sec:db-background}

\subsection{Background}
\begin{figure}[t]
  \centering
  \includegraphics[width=0.95\columnwidth]{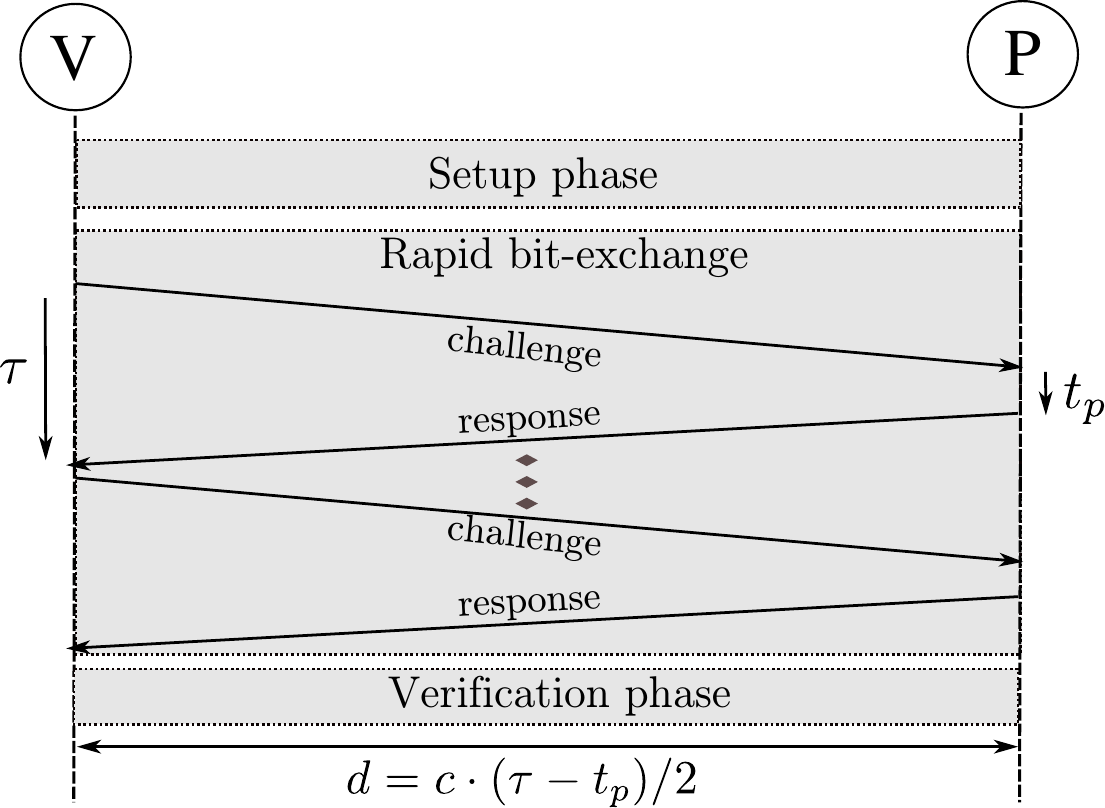}
  \caption{The three phases of a distance bounding protocol. (i) Setup phase
    where specific information gets exchanged between the prover and the
    verifier, (ii) Rapid-bit exchange where single bit challenges and responses
    are exchanged and (iii) Verification phase where the responses are
    validated and distance bound estimated.}
  \label{fig:db-protocol}
\end{figure}

The goal of a distance bounding system is that a verifier establishes an upper
bound on its physical distance to a prover. Distance bounding protocols follow a
specific procedure which typically includes a setup, rapid-bit exchange and
verification phases (Figure~\ref{fig:db-protocol}). In the setup phase, the
verifier and the prover agree or commit to specific information that will be
used in the next protocol phases. In the rapid-bit exchange phase, the verifier
challenges the prover with a number of single-bit challenges to which the prover
replies with single-bit responses. The verifier measures the round-trip times of
these challenge-reply pairs in order to estimate its upper distance bound to the
prover. The distance $d$ between the verifier and the prover is calculated using
the equation $d = \frac{c.(\tau-t_p)}{2}$, where $c$ is the speed of light
($3\cdot10^8\unit{m/s}$), $\tau$ is the round-trip time elapsed and $t_p$ is the
processing delay at the prover before responding to the challenge. The
verification phase is used for confirmation and authentication. It should be
noted that depending on the protocol construction the verification phase may not
be required.

\begin{figure}[t]
  \centering
  \includegraphics[width=0.8\columnwidth]{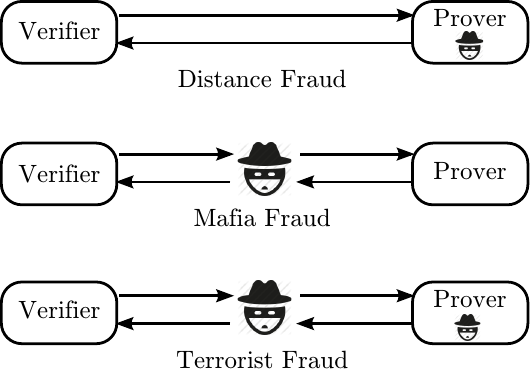}
  \caption{Attacks on distance bounding systems. In distance fraud, an untrusted
    prover tries to cheat on the measured distance. Mafia fraud is achieved by
    an external attacker by relaying information between a trusted prover and
    verifier. In terrorist fraud, the prover colludes with an external attacker
    to cheat on the measured distance.}
  \label{fig:frauds}
\end{figure}

The security of distance bounding protocols is traditionally evaluated by
analyzing their resilience against three types of attacks: \emph{Distance
  Fraud}, \emph{Mafia Fraud} and \emph{Terrorist Fraud} attacks.
Figure~\ref{fig:frauds} illustrates graphically these attack scenarios and the
entities involved. In a \emph{distance fraud attack}, an untrusted prover tries
to shorten the distance measured by the verifier. Since the round-trip time
includes the processing delay, an untrusted prover can reduce the distance
measured by either sending its replies before receiving the challenges or by
computing the responses faster. There is no external attacker involved in this
attack.

\emph{Mafia fraud attacks}, also called relay attacks, were first described by
Desmedt~\cite{DesmedtAug87}. In this type of attack, both the prover and
verifier are honest and trusted. An external attacker attempts to shorten the
distance measured between the prover and the verifier by relaying the
communications between the entities. Distance bounding protocols prevent relay
attacks due to the fact that the time taken to relay the challenges and
responses will only further increase the distance bound estimate. However, it is
important to keep the variance of the prover's processing time to a minimum to
ensure high security guarantees. If the time taken by the prover to process
challenges varies significantly between challenges, the verifier has to account
for the high variance in its distance estimation. Depending on the amount of
variance to be accounted for, an attacker can reduce the distance by relaying
communications between the prover and the verifier.

Finally, in \emph{terrorist fraud attacks}, an untrusted prover collaborates
with an external attacker to convince the verifier that he is closer than he
really is. All countermeasures to terrorist fraud make the assumption that the
untrusted prover does not reveal his long-term (private or secret) key to the
external attacker which he collaborates with.

Recently, another type of attack on distance bounding protocols called the
\emph{distance hijacking} attack was proposed~\cite{CremersMay12}. The authors
give a real world example of a dishonest prover with a stolen smartcard gaining
access to a secure facility; though he is not within the required proximity. The
attacker exploits a honest prover's presence by hijacking its rapid bit-exchange
phase with the verifier. A system's resilience to distance hijacking depends on
the higher level protocol implementation and is independent of the
physical-layer. Therefore, in this work we do not address distance hijacking
attacks.

\subsection{Distance Bounding Implementations}
\label{sec:compare-db-implementations}

A number of distance bounding protocols were proposed following the work of
Brands and Chaum~\cite{BrandsMay93}. These protocols provide resilience against
one or all of the above mentioned attacks. However, the security of these
protocols was mostly analyzed based on information theoretic proofs without
considering physical layer attacks. For example, a protocol is said to be
resilient against distance fraud attacks if the response bits are dependent on
the challenge bits, \ie the prover cannot respond before actually receiving the
challenge. As described previously, a prover's distance is measured based on
some physical layer parameter such as received signal strength or round trip
times. Therefore, in practice, the security of distance bounding protocols also
depends on the actual physical layer design and implementation of the distance
bounding system.

For instance, an untrusted prover can use specialized or modified hardware to
compute a response faster than the delay expected by the verifier to estimate
the distance. It is important to note that a speedup of $1\unit{ns}$ translates
to a distance gain of approximately $15\unit{cm}$. An attacker can also reduce
the distance between the verifier and prover by detecting or demodulating
challenges before receiving them completely or late committing a response as
shown by Clulow et al.~\cite{ClulowSep06}. In order to address these attacks
specific to the physical layer, the focus shifted towards secure physical layer
design of distance bounding systems. Below we summarize the existing
physical layer related designs available in the open literature.
Table~\ref{tab:db-implementations} compares these designs based on the power
requirement, prover processing delay, resilience to distance, mafia, and
terrorist frauds, and the feasibility to implement any distance bounding
protocol.
\begin{table*}\footnotesize
\centering
\begin{threeparttable}[b]
    \newcolumntype{C}[1]{>{\centering\let\newline\\\arraybackslash\hspace{0pt}}m{#1}}
    \begin{tabular}{l|c|c|c|c|c}
    \hline
    \multirow{2}{*}{\textbf{Implementation}}& \multicolumn{3}{c|}{\textbf{Attack Resilience}}&
    \multirow{2}{*}{\textbf{Compatible Protocols}}&\multirow{2}{*}{\textbf{Power Req\tnote{b}}}\\ \cline{2-4}
     &\textbf{DF (processing delay)\tnote{a} } & \textbf{MF} & \textbf{TF} & & \\ \hline
        Tippenhauer~\cite{Tippenhauer12}& $\checkmark$ ($100\unit{ns}$) &
    $\checkmark$ & $\checkmark$ & Any & High\\
    Hancke~\cite{HanckeMay11} & $\checkmark$ ($40\unit{ns}$) &
    $\checkmark$ & $\times$ &
    HKP~\cite{HanckeSep05} & High \\
    CRCS~\cite{RasmussenAug10}\tnote{c} & $\checkmark$ ($~1\unit{ns}$) &
    $\checkmark$  & $\times$ & CRCS & High \\
    Ranganathan~\cite{RanganathanSep12}\tnote{c}& $\checkmark$ ($3\unit{ns}$)
    & $\checkmark$ & $\checkmark$ & HKP
    based~\cite{KimDec08,TuSep07,ReidMar07} & High \\
    Our Work & $\checkmark$ (\tnote{d}\hspace{5pt}) & $\checkmark$ &
    $\checkmark$ & Any & Low \\\hline
    \end{tabular}
    \begin{tablenotes}
    {\small
    \item[a] A $1\unit{ns}$ prover processing delay enables a maximum distance
      reduction of $15\unit{cm}$ by a dishonest prover.    
    \item[b] Power consumption at the prover.
    \item[c] Focused primarily on reducing the prover's processing delay and
      used frequency switching to communicate data. 
    \item[d] The use of slots enables us to decouple the distance estimation
      from the processing delay.
     }
    \end{tablenotes}
\end{threeparttable}

  \caption{Comparison of the existing distance bounding implementations in
    prior art.}
\label{tab:db-implementations}
\end{table*}


Initial distance bounding implementations~\cite{SastrySep03,RasmussenOct08}
proposed the use of both radio frequency and ultrasound. The verifier that wants
to securely verify the location claim of a prover transmits a challenge using RF
and the prover responds back using ultrasound. Based on the time-of-arrival of
the ultrasound packet, the location claim `$l$' of the prover and the
propagation time of radio and ultrasound signals in air, the verifier estimates
the prover's distance `$d$'. If `$d$' is larger than the claimed distance `$l$',
then the verifier rejects the prover's location claim. The authors reasoned out
that the use of RF communication in both directions would make the prover's
processing delay very large making the system unusable. One of the main problems
with these systems is that an untrusted prover or an external attacker with a
proxy node in the verifier's region of interest can take advantage of this. By
using radio frequency as a wormhole channel to echo the response back to the
verifier, the attacker can reduce the round-trip-time and hence the distance
estimate. Hence, it became essential to develop new methods to reduce the
prover's complexity and processing delay.

\paragraph{Hancke's Distance Bounding Channel} Hancke and
Kuhn~\cite{HanckeSep05} introduced one of the first distance bounding protocols
suitable for computationally constrained devices such as RFID with a specific
prover design. Subsequently, Hancke~\cite{HanckeMay11} further extended this
work with a UWB communication channel. In the proposed channel, the verifier
(here, the RFID reader) embeds the challenge bits as ultra-wideband pulses in
addition to the transmitted carrier signal. These pulses are transmitted with a
delay after every rising edge of the carrier signal. This delay is known apriori
to both the verifier and the prover. The presence or absence of the pulse
indicates whether the challenge bit is '$1$' or '$0$'. The prototype
implementation resulted in distance bounds for near field RFID up to $1\unit{m}$
for trusted provers and $11\unit{m}$ in case of untrusted provers. Several
challenges exist in implementing this design. First, since the communication
link includes both low-frequency carrier and the ultra-wideband pulses, the RFID
tag receiver architecture complexity increases drastically. Second, the
ambiguity in distance still depends on the processing delay of the prover.
Hence, an untrusted prover with access to faster hardware can reduce the
processing delay thereby cheating on the distance estimated by the verifier.

\paragraph{Tippenhauer's UWB Distance Bounding System}
Tippenhauer~\cite{Tippenhauer12} designed and implemented a distance bounding
system with focus on optimizing the rapid bit-exchange phase. Due to the ranging
precision and resilience to multipath effects, an impulse radio ultra-wideband
(IR-UWB) physical layer was used for communication. IR-UWB systems communicate
data using short pulses which are typically $2-3\unit{ns}$ long. Range
estimation is based on the time elapsed between transmitting a challenge pulse
and receiving a corresponding response. In any distance bounding protocol the
rapid bit-exchange phase is the core and the final distance estimation is based
on the exact timing of these challenge and response pulses. Since the design
primarily focused on the fast rapid bit-exchange phase, any distance bounding
protocol can be implemented and deployed using this system. The processing delay
at the prover depends on the protocol adopted e.g., the XOR processing function
used in the prototype implementation resulted in an overall delay of $\approx
100\unit{ns}$. However, the narrow IR-UWB pulses utilize a large bandwidth
($>500\unit{MHz}$) which require both the prover and the verifier to be equipped
with high sampling rate ADCs and DACs to receive and transmit IR-UWB pulses
respectively. Such high sampling rate ADCs and DACs consume significant power
(typically around $1-4 \unit{W}$) making it infeasible for applications where power
consumption at the prover needs to be low (order of few mW or
$\mu$W).

\begin{figure*}[t]
  \centering
  \includegraphics[width=0.95\textwidth]{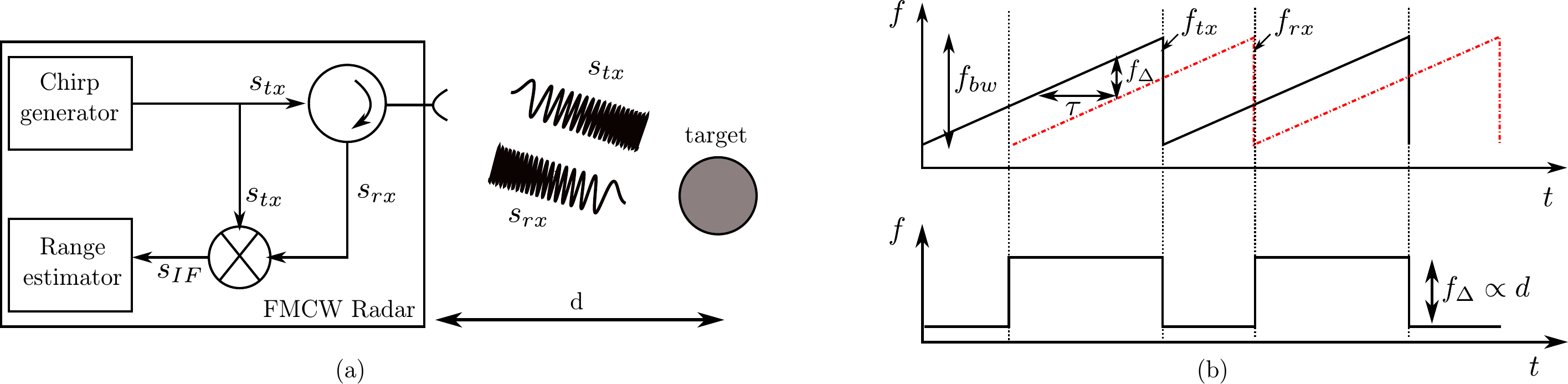}
  \caption{(a) Conventional FMCW-based radar system comprising of a chirp
    generator, mixer and a signal processing block to estimate range. (b)
    Ranging principle: The beat frequency $f_{\Delta}$ is the difference
    between the instantaneous transmit frequency and the frequency of the
    reflected signal. This beat frequency is proportional to the round-trip
    time delay $\tau$ for the signal to be received after being reflected off
    the target object.}
  \label{fig:fmcw-concept}
\end{figure*}

\paragraph{Rasmussen's Challenge Reflection with Channel Selection (CRCS)}
The CRCS~\cite{RasmussenAug10} scheme reduced the prover's processing delay to
$1\unit{ns}$ by eliminating the need for interpreting the challenge during the
rapid-bit exchange phase. In this implementation, the challenges are reflected
back by the prover on different frequency channels. Given that the incoming
challenge is not interpreted during the time-critical phase, the majority of
state-of-art distance bounding protocols (e.g., Brands-Chaum, Hancke-Kuhn)
cannot be realized using this scheme. In addition, the lack of challenge
demodulation made this scheme vulnerable to terrorist fraud attacks. For
example, as shown in~\cite{RanganathanSep12} an untrusted prover can
pre-calculate the responses (since they are independent of the challenge signal
in CRCS) and forward them to a colluding attacker located near the verifier. The
colluding attacker can then successfully execute the rapid-bit exchange phase
with the verifier. Based on the CRCS scheme, Ranganathan et
al.~\cite{RanganathanSep12} proposed a hybrid analog-digital prover design that
is resilient to terrorist fraud attacks with a prover processing delay of
approximately $3\unit{ns}$. This design can be used to implement all distance
bounding protocols that follow the Hancke-Kuhn protocol construction \ie the
response is selected from one or more registers based on the challenge. Both
works focused on minimizing the challenge processing delay at the prover through
architectural modifications with very few details on the physical layer
characteristics of the radio frequency signal which plays a critical role in
ranging precision and data communication. In addition, the absence of challenge
interpretation during the rapid-bit exchange phase makes the system vulnerable
to simple response replay attacks. In order to prevent such attacks, the prover
needs to demodulate, store and communicate the challenges back to the verifier
during the final verification phase of the protocol. This increases the
complexity making it challenging to realize low power provers.

We summarize and compare the aforementioned implementations in
Table~\ref{tab:db-implementations}. With the exception of~\cite{Tippenhauer12},
all the implementations have limitations on the higher layer distance bounding
protocols that can be implemented. In all these systems, the resilience to
distance fraud attacks depends on the processing delay at the prover. For
example, in the case of CRCS~\cite{RasmussenAug10} where the prover has a
processing delay of $1\unit{ns}$, an untrusted prover can cheat on its distance to a
maximum $15\unit{cm}$. Distance bounding protocols inherently protect these
systems against conventional amplify and forward relay attacks. However, their
resilience to a stronger attacker capable of detecting challenges earlier than a
conventional receiver (more details discussed in
Section~\ref{subsec:mafia-fraud}) is dependent on the physical properties of
the transmitted information (\eg symbol duration, type of modulation scheme
used). In addition, the physical layer scheme also affects the ranging
precision, complexity of prover design and therefore its power consumption. The
complex design and strict hardware requirements (\eg ADC and DAC requirements)
makes them unsuitable for power sensitive applications. In this work, we fill
this void by proposing a complete physical layer scheme, specifically designed
for distance bounding that can be leveraged to build low power provers.
 
\section{FMCW based Distance Bounding}
\label{sec:fmcw-based-db}

\subsection{FMCW Basics}
\label{sec:fmcw-radar-background}

Monotone (or single frequency) radars transmit pulses of short duration and
measure distance based on the round-trip time of the received pulse reflected
off the target. Such radars are more susceptible to channel interference. In
Frequency Modulated Continuous Wave (FMCW) radar~\cite{StoveOct92}, chirp
signals~\cite{BerniJun73} are used to determine range and velocity of a target.
Figure~\ref{fig:fmcw-concept}(a) illustrates the basic building blocks of a
conventional FMCW radar system. The radar base station transmits a chirp signal
($s_{tx}(t)$) which gets reflected off the target object back to the base
station. The reflected signal ($s_{rx}(t)$) is then mixed with the transmitted
signal at that instant to produce a ``beat frequency''. The beat frequency
($f_{\Delta}$) is proportional to the round-trip time ($\tau$) taken to receive
the reflected chirp signal; thereby able to measure distance $d$ to the target
object.

The transmitted chirp signal $s_{tx}(t)$ is mathematically represented as
shown below.

\begin{equation}
  \label{eq:s_tx}
  s_{tx}(t) = cos(2\pi f_{tx}(t)t)
\end{equation}

where $f_{tx}(t)$ is the frequency sweep function given by
Equation~\eqref{eq:f_tx} and $f_0$ is the starting value of the frequency
sweep. $k$ is the rate of frequency sweep and is a quotient of the length of
the chirp signal $T$ and the total bandwidth $f_{bw}$ swept \ie $k=f_{BW}/T$.

\begin{equation}
  \label{eq:f_tx}
  f_{tx}(t) = f_0 + kt
\end{equation}

The transmitted chirp is reflected off the target object at distance $d$ and
is received back at the radar base station as $s_{rx}(t)$.

\begin{equation}
  \label{eq:s_rx}
  s_{rx}(t) = cos(2\pi f_{rx}(t)t)
\end{equation}

The frequency of the reflected signal can be represented in terms of the
instantaneous frequency of the transmitted chirp as

\begin{equation}
  \label{eq:9}
  f_{rx}(t) = f_{tx}(t-\tau) = f_o + k(t-\tau)
\end{equation}

Mixing the signals $s_{rx}(t)$ and $s_{tx}(t)$ results in an intermediate
frequency signal $s_{IF}(t)=s_{rx}(t)\cdot s_{tx}(t)$ which consists of
frequency components $f_{tx}(t)+f_{rx}(t)$ and $f_{tx}(t)-f_{rx}(t)$. The
difference component is termed as the ``beat frequency'' given by

\begin{equation}
  \label{eq:f_delta}
  f_{\Delta}=f_{tx}(t)-f_{rx}(t) = f_{tx}(t)-f_{tx}(t-\tau)
\end{equation}

Simplifying and representing $\tau$ in terms of distance $d$, \ie
$d=2\cdot\tau/c$, where $c$ is the speed of light ($3 \cdot 10^8\unit{m/s}$),
distance of the target object from the radar base station is estimated using
Equation~\eqref{eq:for_d}.

\begin{equation}
  \label{eq:f_delta_expand}
  f_{\Delta}=k\tau = \frac{f_{bw}}{T} \cdot \tau
\end{equation}

\begin{equation}
\label{eq:for_d}
 d = \frac{c \cdot f_{\Delta} \cdot T_{s}}{2 \cdot f_{bw}}
\end{equation}

Maximum measurable distance and range resolution are two important performance
metrics of any ranging system. Maximum measurable distance $d_{max}$ is the largest
value of distance $d$ that can be measured using a particular ranging system.
In a FMCW radar, this is dependent on the time duration $T$ of the chirp signal
and is given by $d_{max}=cT$. Range resolution $\delta R$ is the minimum change in
distance that can be detected and is proportional to the time resolution of
$s_{tx}(t)$. In other words, $\delta R$ is inversely proportional to the total
bandwidth swept by the chirp and is mathematically represented as shown in
Equation~\eqref{eq:range_resolution}.

\begin{equation}
  \label{eq:range_resolution}
  \delta R = \frac{c}{2 \cdot f_{bw}}
\end{equation}

\subsection{Data Modulation for Distance Bounding}
\label{subsec:ask-fmcw}

\begin{figure*}[t]
  \centering
  \includegraphics[width=\textwidth]{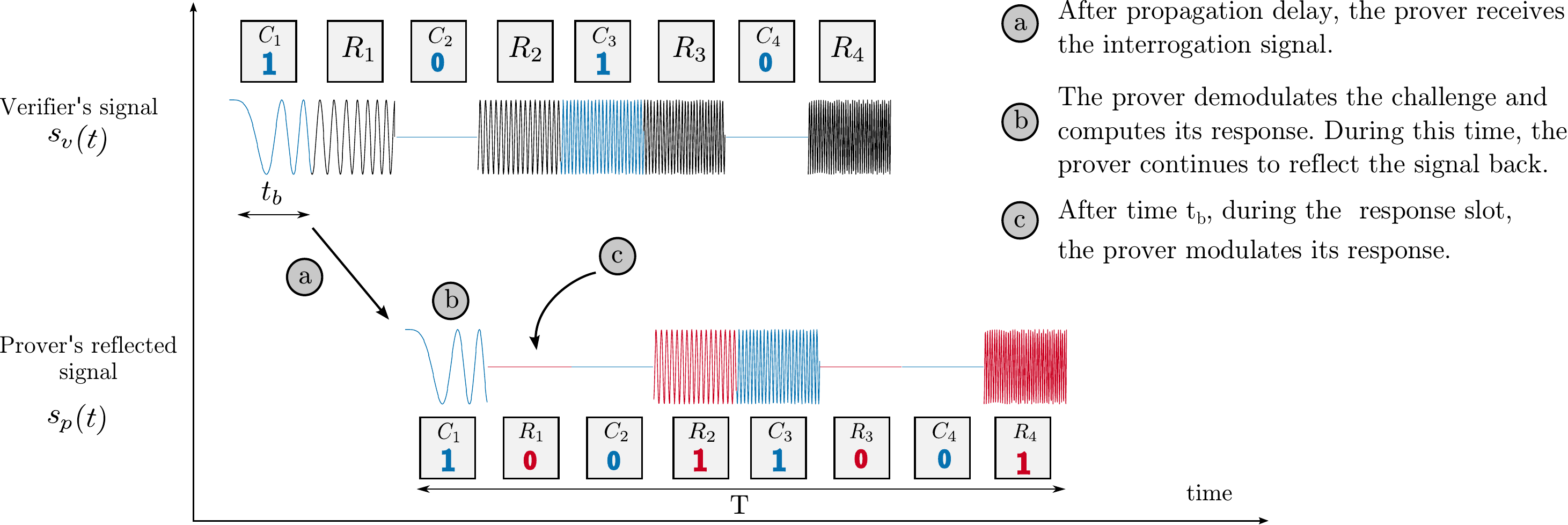}
  \caption{OOK-FMCW packet format: An example OOK-FMCW packet as transmitted by
    the verifier and the corresponding reflected signal from the prover. The
    shown signals are for challenge bits $c[n]=\{1,0,1,0\}$ and the prover's
    processing function is a simple ``invert'' operation. The verifier and
    prover synchronize to these slots using a preamble (not shown in figure).}
  \label{fig:packet-format}
\end{figure*}
\begin{figure*}[t]
  \centering
  \includegraphics[width=0.75\textwidth]{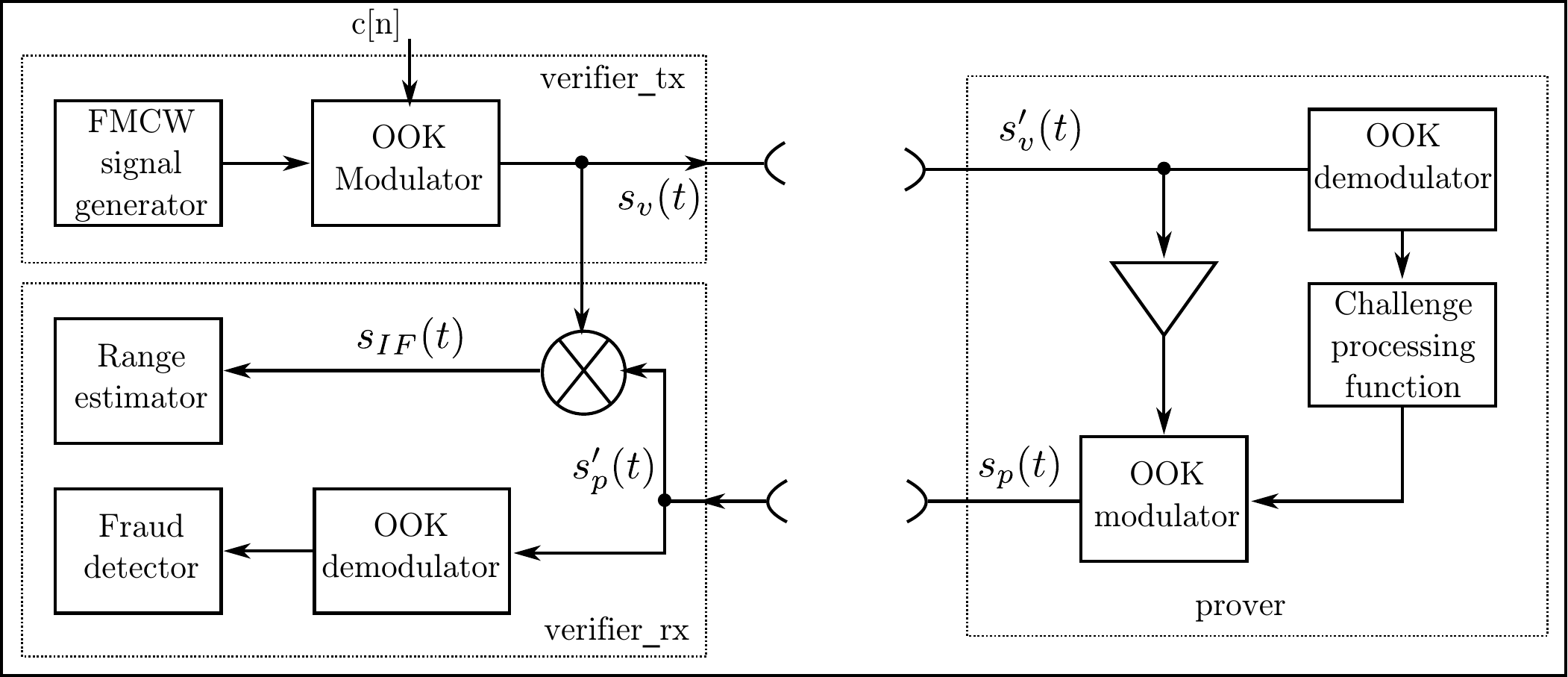}
  \caption{OOK-FMCW based distance bounding system architecture: The
    interrogating signal $s_v(t)$ is an OOK-FMCW transmitted by the verifier.
    The prover receives, demodulates the challenge and computes the response
    while simultaneously reflecting the challenge signal back to the verifier.
    The responses are OOK modulated in the corresponding response time slot. The
    received signal at the verifier is then processed for both range estimation
    and verification of the prover's response.}
  \label{fig:fmcw-db-system}
\end{figure*}

Conventional radar systems do not require any kind of data transmission.
However, in distance bounding protocols the communicating entities (verifier and
prover) exchange challenges and responses during the rapid bit-exchange phase.
This requires data to be modulated over conventional FMCW radar signals. In this
work, we modulate the challenge and response bits over the FMCW chirp signal
using On-Off Keying (OOK). Mathematically, the transmitted signal with OOK
modulation can be represented as

\begin{equation}
  \label{eq:ask-fmcw}
  \sum_{n=1}^Nc[n] \cdot \mathrm{rect}(t-nt_b) s_{tx}(t)
\end{equation}

where $t_b$ is the data-bit period given by $\frac{T}{N}$ ($N$ is the length
of the data packet to be transmitted) and $c[n]$ represents the payload.

The distance bound is estimated similar to conventional FMCW radar systems based
on the ``beat frequency'' $f_{\Delta}$ as shown in Equation~\eqref{eq:for_d}. We
describe the system design in more detail in the next sections.

\subsection{Verifier and Prover Design}
\label{subssec:sys-design}
Figure~\ref{fig:fmcw-db-system} shows the high-level components present in our
system architecture. We focus on the rapid-bit exchange phase since it is,
implementation- and power-wise the most demanding phase of the protocol
execution.

The \emph{verifier's transmitter} ($verifier\_tx$) module consists of an FMCW
signal generator and an OOK modulator. The FMCW signal generator generates a
chirp signal of time duration $T$. The entire chirp signal is divided into
slots, each with time duration $t_b$. The prover synchronizes to these slots
using a preamble that is transmitted by the verifier. The verifier divides the
slots into challenge and reply slots such that every challenge slot is followed
by a response slot. During the challenge slots, the verifier modulates the
challenge bits using OOK modulation and continues to transmit the unmodulated
chirp signal during the response slot (Figure~\ref{fig:packet-format}). The
response slots are used by the prover to transmit its response back to the
verifier. 

When the \emph{prover} receives the challenge signal $s_v^\prime(t)$ from the
verifier, it processes it through two circuits: (i) reflecting and (ii) response
circuits. The reflecting circuit as its name suggests simply reflects the
received signal $s_v^\prime(t)$ after optionally amplifying it (for increased
range). The response circuit is responsible for challenge demodulation and
computation of the prover's response using a processing function. The output of
the processing function is then modulated on top of the reflected signal. We
note that any processing function proposed for distance bounding in prior art
can be used here. Therefore, our proposed physical layer is independent of the
logic-level protocols. The computed response is OOK modulated over the chirp
signal during the corresponding response slot. Like in conventional passive RFID
tags, the prover can simply load modulate its responses back to the verifier. It
is important to note that the prover continues to reflect back the received
signal while simultaneously demodulating the challenges and computing its
response. The propagation delay of the response computation path is one of the
factors that determines the slot duration $t_b$. However, $t_b$ has limited
effect on the system's overall security as explained in
Section~\ref{sec:security-analysis}.

The verifier's receiver module receives the reflected signal $s_p^\prime(t)$
that contains the reflected challenges and the prover's modulated responses and
estimates its distance to the prover. The verifier generates an intermediate
signal $s_{IF}(t)$ by mixing $s_p^\prime(t)$ with $s_v(t)$ as shown in
Figure~\ref{fig:fmcw-db-system} and computes a distance bound by analyzing the
frequency components of $s_{IF}(t)$ as expressed in
Equations~\eqref{eq:f_delta_expand} and~\eqref{eq:for_d}. In addition, the
verifier demodulates and checks the correctness of the prover's responses. It is
important to note that, in a majority of scenarios, the verifier does not have
strict power limitations and therefore the demodulator and signal processing at
the verifier can be implemented as efficiently as possible.

\subsection{Realization of Low Power Provers}
\label{sec:low-power-prover}

\begin{table}
  \centering
  \begin{tabular}[h]{l|l|l}
  \hline
  \textbf{Detector}&\textbf{Operating current}&\textbf{Response time}\\
  \hline
  LTC5536 & $\approx 3\unit{mA}$& $25\unit{ns}$ \\
  AD8313 & $\approx 14\unit{mA}$& $40\unit{ns}$ \\
  AD8314 & $\approx 4.5\unit{mA}$& $70\unit{ns}$ \\
  \hline
  \end{tabular}
  \caption{Operating current values of alternative COTS energy detectors that
    operate in the $0.7-6\unit{GHz}$ frequency range with response times under
    $100\unit{ns}$. All the detectors require a DC voltage bias of $\approx
    3\unit{V}$.} 
  \label{tab:current-ratings-ed}
\end{table}

RFID technology has become ubiquitous in a number of security-critical ranging
applications (\eg commodity goods identification and tracking, physical access
control, automatic toll collection and electronic payment systems). Prior
works~\cite{FrancillonFeb11,FrancisDec10,RolandSep12} showed the vulnerability
of RFID based proximity systems to simple mafia fraud (relay) attacks. One of
the main challenges in enabling distance bounding protocols for these
applications is the tag's strict power constraints. Passive RFID tags do not
have any built-in power sources and derive power by rectifying the received
interrogating signal from the reader. As a result, they are less complex, work
only at short ranges and are incapable of transmitting data on their own.
Passive tags communicate with the reader by modifying the signal received from
the reader. Semi-passive tags have a built-in power source to, for example,
amplify the response signal, but still cannot transmit data independently
without the presence of a reader's interrogating signal.

The proposed FMCW based physical layer scheme would enable realization of
distance bounding systems, with low power consumption at the tag (prover). Our
prover design in Section~\ref{subssec:sys-design} can be implemented in passive
and semi-passive RFID tags operating in the ISM 2.4~GHz and 5.8~GHz bands using
80~MHz and 150~MHz~\footnote{In theory, 80 MHz gives distance resolution of 1.87
  m, 150 MHz of 99 cm} bandwidth respectively to achieve high distance
precision. Since our system targets short-range distance measurement
applications (less than $5\unit{m}$), the use of $6-8.5\unit{GHz}$
spectrum~\cite{hirt2007european} is also possible. Given that these tags (\eg
~\cite{dardari2008passive,d2012uwb,seetharam2007battery}) already have
backscatter communication capability to send back the distance bounding
response, the only addition would be to incorporate the response computing
function which can be as simple as an inverter or an XOR operation. There are
already several commercially available radio frequency energy detectors that
operate in the above mentioned frequency bands with integrated comparators and
amplifiers. In addition, the response time of these detectors are well under
$100\unit{ns}$ and consume less than $15\unit{mA}$ of current.
Table~\ref{tab:current-ratings-ed} lists a few commercially available detectors
with the above mentioned specifications that can be integrated into state-of-art
RFID tags for an additional power consumption of $\approx 10\unit{mW}$.

Furthermore, it should be noted that passive FMCW-based RFID tags have already
been deployed for asset localization in industrial settings~\cite{symeo}.
In-order to increase the maximum range that can be measured, FMCW-based
semi-passive RFID tag designs were also explored. For example, the pulsed
reflector design of~\cite{WehrliFeb10} can measure distances with a ranging
precision of $15-30\unit{cm}$ and a low power consumption of $~54\unit{mW}$.
In~\cite{StrobelSep11}, the authors present a circuit design for BiCMOS
integrated circuits with a power consumption of $150\unit{mW}$. Thus, using our
proposed physical layer scheme it is indeed possible to realize provers that
consume low power suitable for deployment in power-constrained environments.

\section{Security Analysis}
\label{sec:security-analysis}

In this section, we analyze the security of our proposed system under the
distance, mafia and terrorist fraud attacks.

\begin{figure}[t]
  \centering
  \includegraphics[width=\columnwidth]{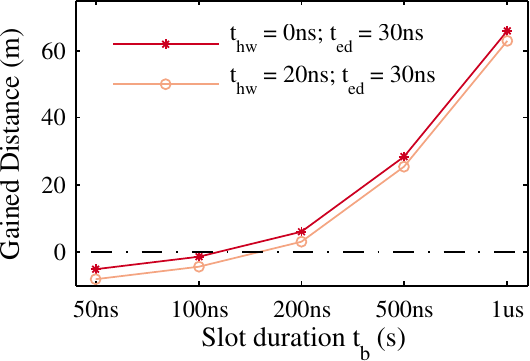}
  \caption{Maximum distance an attacker can cheat by performing an
    early-detect and late-commit attack on the physical layer of the symbol.}
  \label{fig:mafia_fraud_plot}
\end{figure}

\subsection{Distance Fraud}
\label{subsec:distance-fraud}
In a distance fraud, an untrusted prover claims to be at a distance closer than
the actual one. In conventional secure ranging systems, an untrusted prover can
shorten the measured distance either by modifying its internal processing delay
time or by replying before receiving the complete challenge signal. In the
former, the prover implements an improved hardware to process the challenges
faster than the ``processing delay'' accounted in the distance estimation at the
verifier. In the latter case, the prover early detects the challenge signal,
computes and transmits back the response (sometimes later than required also
referred to as ``late commit''~\cite{ClulowSep06}) resulting in faster
processing thereby reducing the distance estimated by the verifier.

In our system, the dishonest prover does not
gain any distance advantage by speeding up response computation as distance is
estimated solely based on the beat frequency created by mixing the reflected
signal with the transmitted FMCW signal. The slot assignment to challenge and
response bits forces the prover to wait until the challenge is reflected before
modulating the response on the response slot. Early modulation would corrupt the
challenge signal thereby being detected at the verifier during the response
validation phase. Also, the prover does not gain any distance by executing such
an early response attack as the distance estimation based on FMCW is completely
decoupled from the data response at the prover.

The same reasoning holds for an untrusted prover who early detects the challenge
signal, computes and late commits the response without any colluding entity in
close proximity to the verifier. Irrespective of how fast the prover detects and
processes the challenge, unless the prover reflects the signal from close
proximity to the verifier, he will not be able to cheat on the measured
distance.

\subsection{Mafia Fraud}
\label{subsec:mafia-fraud}

Mafia fraud attacks are also called relay attacks and were first described by
Desmedt~\cite{DesmedtAug87}. The attacker is an external entity who attempts to
shorten the distance estimated by the verifier by relaying communications
between the verifier and the honest prover. There are two ways in which an
attacker can carry out a mafia fraud at the physical layer: (i) Amplify and
forward (ii) Early-detect and late commit of data symbols.

\emph{Amplify and forward:} In this method, the attacker simply amplifies and
relays communication between the verifier and the prover. The attacker does
not modify any physical layer characteristic of the symbol. Such a method is
insufficient for an attacker since the effective distance computed would still
be the actual distance between the trusted prover and the verifier.

\emph{Early-detect and late-commit:} Clulow et al.~\cite{ClulowSep06} introduced
the early-detect and late-commit attacks where a successful attacker early
detects (ED) the symbols from the verifier and late commits (LC) those signals
from the prover back to the verifier. The feasibility of ED and LC attacks on
RFID was demonstrated in~\cite{HanckeApr08}. Here, we analyze the resilience of
the proposed OOK-FMCW physical layer against ED and LC attacks. In order to
successfully execute the attack, the attacker must do the following: (i)
early-detect the challenge from the verifier, (ii) communicate it to prover,
(iii) early-detect the response from the prover and finally (iv) late commit a
value back to the verifier. For the analysis, lets consider one challenge and
response slot. Assuming that the verifier requires at least
$50\%$\footnote{Assuming an energy detection based demodulator at the verifier
  with the threshold set to half the maximum symbol energy. This can vary
  depending on the type of receiver used to demodulate data.} of the symbol to
demodulate correctly, an attacker has $t_b + 0.5t_b$ time to respond. Within
this time, the attacker must perform the above mentioned operations. If $t_{ed}$
is the time necessary for the attacker to reliably early-detect the challenge
from the verifier and the response from the prover, $t_{hw}$ is the delay at the
attacker hardware for amplifying and relaying, the time remaining for the
attacker to relay communications is given by,

\begin{equation}
  \label{eq:mafia_fraud_wo_proverdelay}
  t_{mafia} = 1.5t_b - 2t_{ed} - t_{hw}
\end{equation}

Since the prover is trusted, the response will be available only after the
challenge slot time period \ie $t_b$. Therefore,

\begin{equation}
  \label{eq:mafia_fraud_final}
  t_{mafia} = 0.5t_b - 2t_{ed} - t_{hw}
\end{equation}

Therefore the maximum distance an attacker can cheat on can be expressed as,

\begin{equation}
  \label{eq:mafia_fraud_disteq}
  d_{gain}= \frac{c}{2} \cdot (0.5t_b - 2t_{ed} - t_{hw})
\end{equation}

It is important to note that Equation~\eqref{eq:mafia_fraud_disteq} holds good
even in the scenario where an external attacker (in close proximity to the
verifier) reflects the challenge signal back to the verifier resulting in a beat
frequency corresponding to the attacker's distance from the verifier. However,
for a successful attack, the attacker still has to modulate the response after
the challenge slot period $t_b$. This time constraint forces the attacker to
early detect, relay and late commit the challenge and response bits as described
previously and hence the maximum distance gained remains unchanged.

In Figure~\ref{fig:mafia_fraud_plot}, we give an intuition by substituting
nominal values for $t_{ed}$ and $t_{hw}$. In a real world scenario, the values
will depend on various characteristics of the attacker hardware (\eg filter
order, ADC delays, signal group delay, algorithm used to early-detect \etc).
Since $t_b$ is selected based on the delay of the challenge processing
function at the prover, it can be observed that even for an attacker with
ideal hardware ($t_{hw}=0\unit{ns}$), it is impossible to reduce the distance
against a system with prover processing delay of $100\unit{ns}$. 

\begin{figure}[t]
  \centering
  \includegraphics[width=0.75\columnwidth]{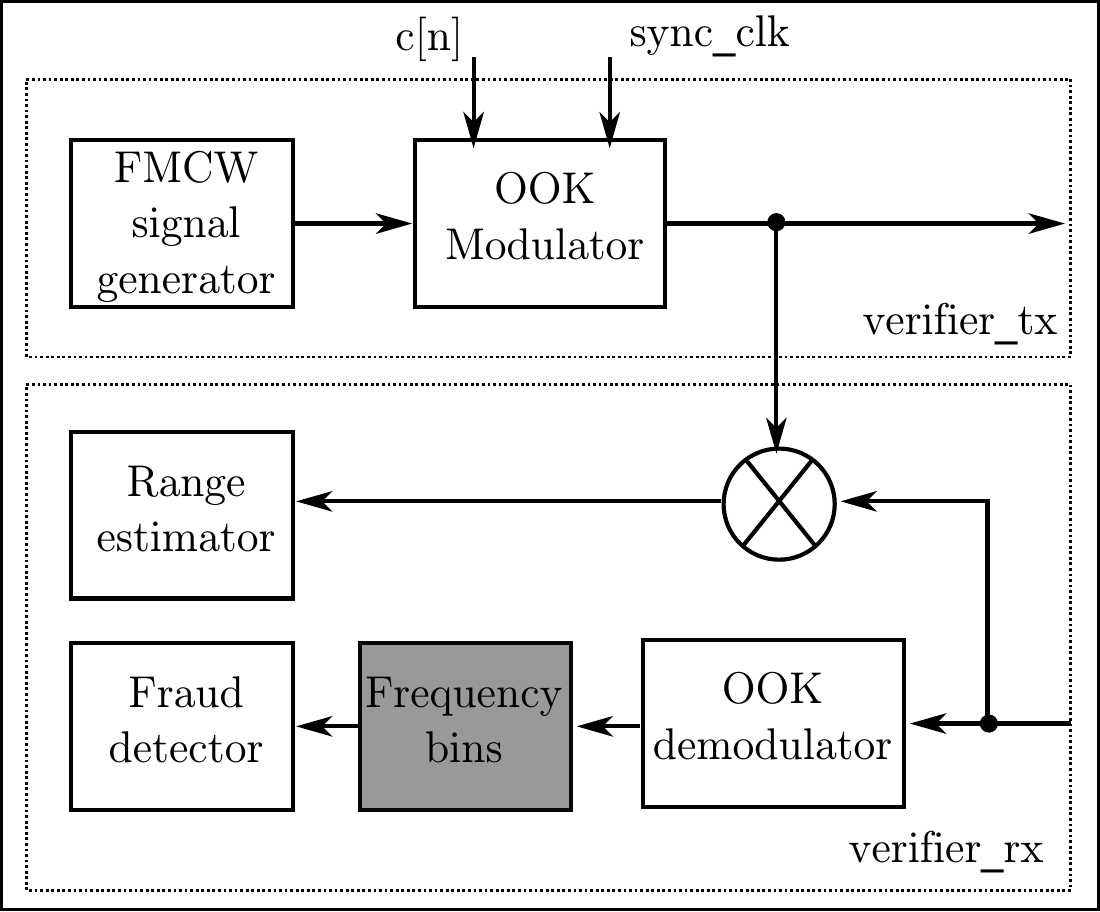}
  \caption{Improved verifier design including the frequency bin based
    late-commit mafia fraud detector.}
  \label{fig:verifier-bins}
\end{figure}

\emph{Mafia fraud detector:}

The linearly increasing frequency characteristic of the chirp signal makes it
feasible to detect mafia fraud attacks by analyzing the frequency components at
specific time intervals. This temporal knowledge of the signal enables us to assign
every challenge and response to one or more frequency bins. Each frequency bin
contains spectral energy values for a range of contiguous frequencies. Specifically,
it is possible to estimate the range of frequencies a particular challenge or
response bit will occupy given a slot period $t_b$, starting sweep frequency $f_0$
and chirp duration $T$. We divide each challenge and response slot into $N$
frequency bins. For a successful attack, the attacker must ED and LC every challenge
and response. A late commit on a symbol would result in incorrect bin values and
this would appear consistently throughout the chirp sweep bandwidth. Thus, by
analyzing the frequency bins for expected spectral energy values, a late commit
attack can be detected. It is safe to assume that the possibility of incorrect
spectral values consistently at specific frequency intervals due to just channel
fading effects is negligible, and such an effect could have occurred due to a LC
attack. The late commit detection can be improved by dividing dividing each slot
into more frequency bins \ie increasing $N$. The modified verifier with the
frequency bin based mafia fraud detection module is shown in
Figure~\ref{fig:verifier-bins}.

\subsection{Terrorist Fraud}
\label{subsec:terrorist-fraud}

In Terrorist fraud attacks, a dishonest prover collaborates with an external
attacker to convince the verifier that he is closer than he really is. The
prover will help the attacker with information as long as it does not reveal
the prover's long term secret. Terrorist fraud resilient
protocols~\cite{TuSep07,KimDec08,ReidMar07} bind the prover's long term secret
to the nonces that are exchanged in the protocol. This prevents the prover
from revealing the nonces to the attacker without disclosing its long term
secret. Since our proposed physical layer is independent of the high-level
protocol, the system security depends on the distance bounding protocol
implemented above the physical layer.

\emph{Special case of terrorist fraud:} Consider the scenario where a nearby
external attacker simply reflects the interrogating OOK-FMCW signals back to the
verifier, while simultaneously relaying the signals to the distant prover. The
untrusted prover colludes with the attacker and helps him authenticate (by
providing the responses) without revealing its long term secret key. The tasks
needed to be executed by the external attacker and the untrusted prover is
similar to that of a mafia fraud attacker as described in
Section~\ref{subsec:mafia-fraud}. However in this setting, the prover colludes
with the attacker and therefore communicates the response as soon as possible.
Thus, the attacker is not constrained by the additional time $t_b$
(Equation~\eqref{eq:mafia_fraud_final}) and the maximum possible distance that
the attacker can cheat is same as that expressed in
Equation~\eqref{eq:mafia_fraud_wo_proverdelay}.

\begin{figure}[t]
  \centering
  \includegraphics[width=\columnwidth]{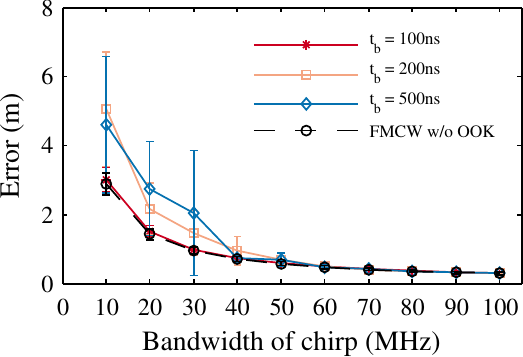}
  \caption{Measurement precision: The mean error in distance estimation against
    bandwidth of the FMCW signal for various slot durations $t_b$. The SNR was
    fixed at $15\unit{dB}$ and the error is a mean value obtained by measuring
    100 different distances within the possible maximum measurable distance.}
  \label{fig:dist_bw_plot}
\end{figure}

\section{System Evaluation}
\label{sec:system-evaluation}

In this section, we evaluate our proposed distance bounding system using both
simulations and experiments. Through simulations, we analyze the bit error rate
and ranging precision due to the on-off keying over FMCW. Then, we
experimentally validate our prover's processing delay and ranging precision
using a prototype.

\subsection{Simulation Model and Analysis}
\label{sec:simulation-setup}

The preliminary analysis through simulations were done using Matlab. The
OOK-FMCW signal is generated by mixing a binary data signal with a chirp. The
duration of a single chirp ($T$) was fixed at $10\unit{\mu s}$ with the initial
sweep frequency $f_0$ set to $2.4\unit{GHz}$. The physical layer parameters such
as the chirp bandwidth $f_{bw}$ and bit-period (duration of each slot) $t_b$ is
made configurable based on the analysis performed. The generated OOK signal is
passed through an additive white Gaussian noise (AWGN) channel. The signal to
noise ratio (SNR) of the channel is varied depending on the analysis performed.
We model the receiver as two submodules: (i) Energy detector for demodulating
data sent over OOK-FMCW and (ii) FMCW-based distance measurement module. For the
energy detection, the threshold value to distinguish the bits `0's and `1's is
set at a value $6\unit{dB}$ lower than the maximum energy estimated for a `1'
bit under no noise conditions. The signal processing for distance estimation is
implemented following the theory described in
Section~\ref{sec:fmcw-radar-background}.

\begin{figure}[t]
  \centering
  \includegraphics[width=\columnwidth]{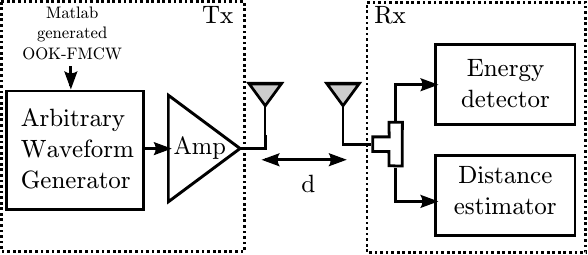}
  \caption{Block level overview of the experimental setup comprising of the
    transmitter and receiver modules.}
  \label{fig:prototype-setup}
\end{figure}

\begin{figure*}[t]
  \centering \subfigure[]{
    \includegraphics[width=0.47\textwidth]{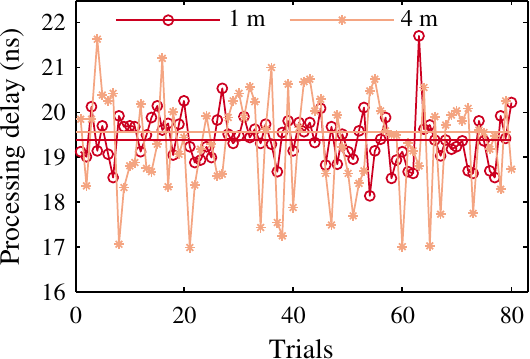}
    \label{fig:exp-variance-plot}
}
  \hspace{0.02\textwidth} \subfigure[]{
    \includegraphics[width=0.46\textwidth]{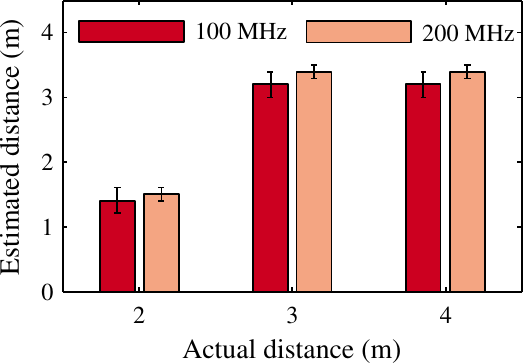}
    \label{fig:exp-range-precision}
}
 
\caption{(a) Challenge processing delays. The median value of $t_p$ was
  approximately $19.5\unit{ns}$ for both the values of $d=\{1\unit{m},
  4\unit{m}\}$. (b) Ranging precision. For $d=\{2,3,4\}\unit{m}$, the errors in
  the estimated distances were less than a meter.}
  \label{fig:exp-results}
\end{figure*}

\paragraph{BER and Ranging Precision}
\label{sec:ber-ranging-precision}
First, we determine the minimum SNR required to reliably communicate data \ie
challenges and responses with the proposed physical layer scheme. In our
simulations we vary the SNR from $0$--$10\unit{dB}$ keeping the slot length
$t_b=100\unit{ns}$ a constant. It is observed that for SNR greater than
$8\unit{dB}$, we were able to demodulate the bits with a BER of $10^{-7}$. Next,
we analyze the effect on ranging precision due to the OOK modulation over
conventional FMCW radar. In addition to $T$, SNR is set to a constant
$15\unit{dB}$. For a specific $t_b$, the error in distance measured is
determined for various values of $f_{bw}$. The error is a mean value obtained by
measuring $100$ different distances within the possible maximum measurable
distance $d_{max}$. The simulations are repeated for
$t_b=\{100\unit{ns},200\unit{ns},500\unit{ns}\}$ and the results are shown in
Figure~\ref{fig:dist_bw_plot}. It is observed that the challenge slot period
$t_b$ has limited effect on the distance measurement precision for signals with
bandwidth greater than $50\unit{MHz}$. We note that, even at lower bandwidths,
the observed precision would still be suitable for a wide range of ranging
applications. Alternatively, we could use amplitude shift keying e.g., a signal
with low amplitude can represent a '0' bit as against absence of the signal
itself (as in OOK). We use the above results of our preliminary simulations to
build and evaluate our prover through real experiments.

\subsection{Experimental Setup}
\label{sec:experimental-setup}

\begin{figure}[t]
  \centering
  \includegraphics[width=0.99\columnwidth]{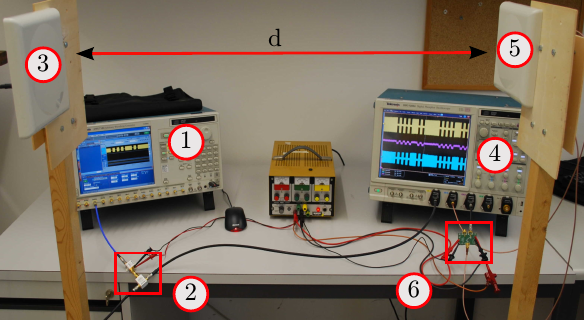}
  \caption{Experimental setup. An arbitrary waveform generator (1) outputs the
    OOK-FMCW samples. The signal is amplified (2) and a part of it is transmitted using a
    planar antenna (3) and the other recorded for distance estimation using a
    storage oscilloscope (4). The received signal (5) is input to the OOK
    detection and comparator circuit (6) and to the storage oscilloscope.}
  \label{fig:exp-photo}
\end{figure}

In this section, we describe the experimental setup
(Figure~\ref{fig:prototype-setup}) used to evaluate our proposed distance
bounding system. Our experiments primarily focuses on the two critical
parameters of any distance bounding system: (i) Challenge processing delay and
(ii) Ranging precision.

A picture of our experimental setup is shown in Figure~\ref{fig:exp-photo}. The
transmitter consists of an arbitrary waveform generator (AWG) capable of
generating signals at a sampling rate of $20\unit{GSa/s}$, a $20\unit{dB}$ radio
frequency amplifier and a directional planar antenna. The OOK-FMCW signals are
generated using Matlab as described in Section~\ref{sec:simulation-setup} and
loaded into the AWG. The OOK-FMCW signals are amplified and transmitted using a
planar antenna. At the receiver, the signals are captured using a planar antenna
similar to the one used at the transmitter. The received signal is recorded on a
$50\unit{Gsa/s}$ digital storage oscilloscope (DSO). In addition, the received
signal is input to the challenge demodulator circuit~\cite{ltc5564datasheet}
which essentially is a Schottky RF peak detector with programmable gain and a
high speed comparator. The output of the demodulator circuit is also observed on
the oscilloscope. We evaluate our system for different configurations of
OOK-FMCW signals with the initial sweep frequency $f_0$ set to $2.4\unit{GHz}$.
The various physical characteristics of the signals used in the evaluations are
listed in Table~\ref{tab:exp-sig-config}.

\begin{table}
  \centering
  \begin{tabular}[h]{l|l}
  \hline
  \textbf{Parameter}&\textbf{Value}\\
  \hline
  Sweep bandwidth $f_{bw}$ & $100$, $200\unit{MHz}$\\
  Slot period $t_b$ & $100$, $250\unit{ns}$\\
  Modulation index & $75$, $100\unit{\%}$\\
  \hline
  \end{tabular}
  \caption{Different configurations of the signals used in the experimental analysis.}
  \label{tab:exp-sig-config}
\end{table}

\subsection{Experimental Results}
\label{sec:experimental-results}

\textbf{Challenge Processing Delay $t_p$:} The challenge processing delay $t_p$
plays an important role in deciding the duration of the challenge and response
slots $t_b$. In our experimental setup, $t_p$ is the time delay for the energy
detector to demodulate the received OOK-FMCW challenge signal and switch the
output of the comparator. For accurate time delay measurements, the signals are
pre-processed by applying Hilbert transform and passing it through a median
filter (to preserve the rising and falling edges while reducing noise).
Figure~\ref{fig:exp-variance-plot} shows the response times observed at the
receiver over a number of trials. The processing delay was measured with the
receiver placed at $1$ and $4\unit{m}$ away from the transmitter. The medial
delay observed was about $19.5\unit{ns}$ and remained largely unaffected due to
distance from the transmitter. Hence, the value of $t_b$ can be further reduced
to about $50\unit{ns}$ (including fall-time) without affecting the decoding of
challenge bits. Additionally, it is observed that the $t_p$ values show greater
variance with distance due to the variations in the received signal's energy
between trials.

\textbf{Ranging Precision:} In order to evaluate the ranging precision, we
placed the receiver at distances $2$, $3$ and $4\unit{m}$ from the transmitter.
The distance bound is calculated using standard FMCW techniques as described in
Section~\ref{sec:fmcw-radar-background} and the results are plotted in
Figure~\ref{fig:exp-range-precision}. It can be observed that our prototype has
a ranging precision of less than a meter for the evaluated short distances. Due
to the limitations of our experimental setup, we could not measure longer
distances. A combination of factors such as range resolution $\delta R$ (and
hence signal bandwidth), channel multipaths and the receiver sampling rate
affect the precision of a ranging system. Other physical characteristics of the
OOK-FMCW signal such as modulation index, bit (slot) period $t_b$ and duration
of chirp $T$ had no effect on the precision of the ranging system.

\section{Conclusion}
\label{sec:conclusion}

In this work, we proposed and analyzed a new physical layer scheme designed
specifically for enabling distance bounding for short-range, low-power
application scenarios. In this proposal, we combined on-off keying and frequency
modulated continuous wave to design a prover that can potentially be integrated
into passive and semi-passive RFID tags. OOK-FMCW guarantees the distance bound
independent of the processing delay at the prover; irrespective of the distance
bounding protocol implemented. Through our security analysis, we showed that our
system is resilient against distance, mafia and terrorist fraud attacks. For
slot durations less than $100\unit{ns}$, we showed that our system is fully
resilient against an attacker capable of early detection and late commit of the
challenge and response bits. We experimentally validated our distance bounding
system's ranging precision and challenge processing delay. In addition, our
experiments reveal that it is indeed feasible to realize low-power provers that
can process challenges as fast as $\approx 20\unit{ns}$.

As part of future work, we intend to build a complete prototype to fully
evaluate our system's power and performance characteristics. In addition, the
feasibility of using other modulation methods (\eg ASK, PSK) over FMCW remains
to be explored. The nominal values for the security relevant parameters such as
the time required to early detect and late commit under these modulation schemes
also needs to be investigated further.

\bibliographystyle{acm}
\bibliography{aanjhan}

\end{document}